\documentclass[referee]{raa}            

\usepackage{graphicx,times}            
\usepackage{natbib}
\usepackage{lineno}
\usepackage{subfigure}
\usepackage{epstopdf}
\usepackage{caption}
\usepackage{latexsym}
\usepackage{overpic}
\begin{document}


\title{Calibration and performance of the neutron detector onboard of the DAMPE mission
}

   \volnopage{Vol.0 (200x) No.0, 000--000}      
   \setcounter{page}{1}          

   \author{Yong-Yi Huang\inst{1, 2}, Tao Ma\inst{1}\footnote{Corresponding Author: matao@pmo.ac.cn}, Chuan Yue\inst{1}\footnote{Corresponding Author: yuechuan@pmo.ac.cn}, Yan Zhang\inst{1}, Ming-Sheng Cai\inst{1}, Jin Chang\inst{1}, Tie-Kuang Dong\inst{1}, Yong-Qiang Zhang\inst{1}
   }

   \institute{Key Laboratory of Dark Matter and Space Astronomy, Purple Mountain Observatory, Chinese Academy of Sciences, 10 Yuanhua Road, Nanjing 210034, China\\
        \and
        University of Chinese Academy of Sciences, Beijing 100049, China
   }

   \date{Received~~2009 month day; accepted~~2009~~month day}

\abstract{ The DArk Matter Particle Explorer (DAMPE), one of the four space-based scientific missions within the framework of the Strategic Pioneer Program on Space Science of the Chinese Academy of Sciences, has been successfully launched on Dec. 17th 2015 from Jiuquan launch center. One of the most important scientific goals of DAMPE is to search for the evidence of dark matter indirectly by measuring the spectrum of high energy cosmic-ray electrons. The neutron detector, one of the four sub-payloads of DAMPE, is designed to distinguish high energy electrons from hadron background by measuring the secondary neutrons produced in the shower. In this paper, a comprehensive introduction of the neutron detector is presented, including the design, the calibration and the performance. The analysis with simulated data and flight data indicates a powerful proton rejection capability of the neutron detector, which plays an essential role for TeV electron identification of DAMPE.
\keywords{Neutron Detector, Particle Identification, Calibration, Simulation}
}

  \authorrunning{  }            
  \titlerunning{  }  

   \maketitle

%
%
\section{Introduction}           
Over the past few years, a few balloon-borne and space-borne experiments have published exciting results about dark matter indirect detection by measuring the spectra of cosmic-ray electrons and positrons [\citealt{ChangJ2008, Adriani2011, Aguilar2014}]. The DArk Matter Particle Explorer (DAMPE), a satellite-borne high energy particle detector supported by the strategic priority science and technology projects in space science of the Chinese Academy of Science [\citealt{ChangJ2014}], has been successfully launched into a sun-synchronous orbit at the altitude of 500 km on Dec.17th 2015 from Jiuquan launch center [\citealt{ChangJ2017}]. With an unprecedented energy resolution of 1.5\% at 800 GeV [\citealt{ZhangZ2016}], DAMPE offers an excellent potential for dark matter indirect detection by measuring electrons\footnote{As DAMPE cannot discriminate between electrons and positrons, electrons represent electron/positrons in this paper} and gamma-rays in a large energy range from 5 GeV to 10 TeV[\citealt{Ambrosi2017, YuanQ2018, PanX2018}]. Apart from dark matter indirect detection, DAMPE also provides a new opportunity for advancing our knowledge of cosmic ray physics[\citealt{AnQ2019, YueC2020}] and gamma-ray astronomy[\citealt{XuZ2018}].

 \begin{figure}[!ht]
   \centering
   \includegraphics[width=0.6\textwidth, angle=0]{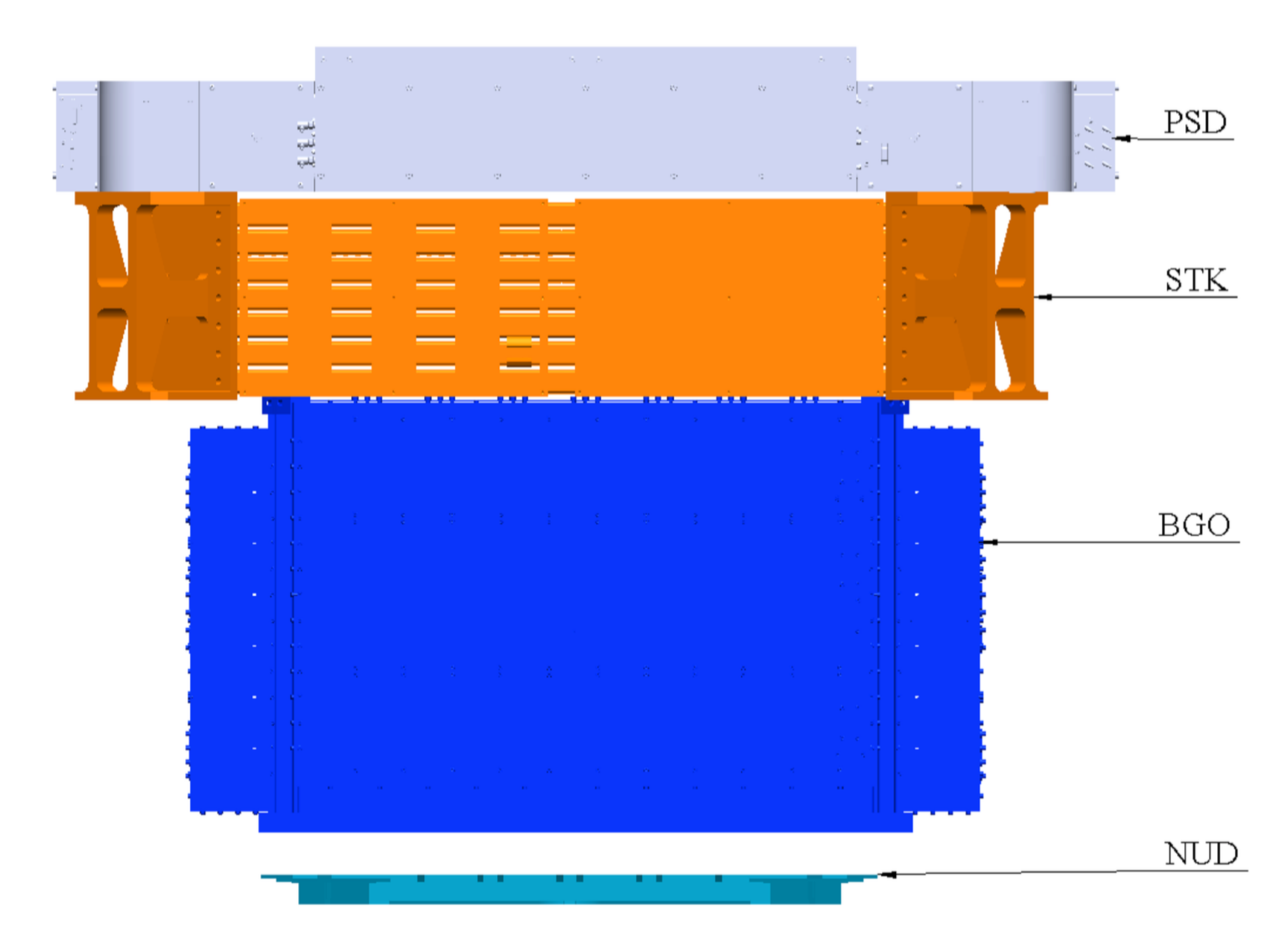}
   \caption{The schematic view of DAMPE}
   \label{fig1}
   \end{figure}

The scientific payload of DAMPE, from top to bottom, consists of four sub-detectors including a Plastic Scintillator strip Detector (PSD) [\citealt{YuY2017, DingM2019}], a Silicon-Tungsten tracKer-converter (STK) [\citealt{Azzarello2016}], a Bismuth- Germanium Oxide (BGO) imaging Energy CALorimeter (ECAL) [\citealt{ZhangY2012}], and a NeUtron Detector (NUD) [\citealt{HeM2016}]. The layout of DAMPE payload is shown in Figure~\ref{fig1}. The PSD is used to measure the particle charge up to $Z=28$ [\citealt{DongT2019}] and identify gamma-rays from charged particles; The STK is used to measure the charges and trajectories of incident particle and reconstruct the directions of incident photons converting into $e^{+}+e^{-}$ pairs; The BGO-ECAL is used to measure the energy deposit of incident particle and provide the electron/proton (e/p) separation based on the shower morphology; The NUD is used to improve the e/p separation power by detecting the neutrons generated by the shower in BGO-ECAL. The hadronic shower would produce a large number of secondary neutrons through the interaction of hadrons (mainly protons) with the detector material. By detecting the secondary neutrons, NUD provides a powerful capability to distinguish electromagnetic showers from hadronic ones, especially in high energy range above TeV.
However, primarily due to the imperfection of NUD simulation at that moment, the signals of NUD were not used in the electron spectrum analysis published in 2017 [\citealt{Ambrosi2017}].In this paper, a comprehensive overview of the NUD is presented, including the design, the calibration and the performance in orbit.

\section{Design of the neutron detector}

\subsection{Principle}

The BGO-ECAL can effectively separate the electromagnetic shower induced by an electron and the hadronic shower induced by a proton by imaging the shower morphology. However, with the increase of energy, the BGO-ECAL alone performance is insufficient for satisfying the proton rejection requirement. In this case, a neutron detector is designed to provide additional e/p separation power in high energy range. The purpose of NUD is to further distinguish electrons from protons by detecting the secondary neutrons from the hadronic shower in the BGO-ECAL. In fact,for a given initial particle of the same energy, the neutron content of a hadronic shower is expected to be one order of magnitude larger than that of an eletromagnetic shower. Once the neutron are created, they are quickly thermalized in the BGO-ECAL, and the total neutron activity over a few microseconds is measured by NUD[\citealt{ChangJ2017}].

The neutron detector is composed of four identical 30cm$\times$ 30cm$\times$ 1.0cm blocks of boron-loaded plastic scintillator (Eljen Technologies EJ-254), EJ-254 is blue-emitting plastic scintillator which contains natural boron at concentrations up to 5\% by weigh, its principal applications are fast neutron spectrometry and thermal neutron detection, the primary function of the boron is to provide a unique scintillation signal for low energy neutrons. The secondary neutrons produced in the BGO-ECAL will be moderated through the elastic collision with the hydrogen atom in the detector material. With a rapid energy decay through the moderation process, some neutrons are moderated into thermal energy range, then the thermal neutrons can be captured by $^{10}$B atoms with the nuclear reaction: n + $^{10}$B $\rightarrow$ $^{7}$Li + $\alpha$ + $\gamma$. The $\alpha$  particles produced by the nuclear reaction will deposit energy through ionization, and generate fluorescent photons in the plastic scintillator. These fluorescent photons are then collected by the photomultiplier (PMT) and converted into electronics signals.  To eliminate the ionization signals from the secondary charged particles within $\sim2\mu$s after the shower development, we set a time-delay gate of 2.5 $\mu$s after the event trigger in orbit [\citealt{Ambrosi2019}]. We thus record only the signals mainly come from secondary neutrons.

\subsection{Requirements}

To achieve good performance for detection, the neutron detector requires a high dynamic range of  2 MeV to 60 MeV, an energy resolution better than 25\%@30MeV and an uniformity of each detection unit less than 25\%.
As a space device, there are some other constraints on the design of NUD: the effective area of the NUD should cover $600 \times 600$ mm$^{2}$, which is the active area of the BGO-ECAL; the envelope size of the NUD should be less than $700 \times 700$ mm$^{2}$; the total weight should be less than 12.5 kg. The whole system of the NUD must keep a high level of performance and stability in the harsh space environment throughout the lifecycle.The mechanical design and thermal design ensure the NUD can survive the mechanics experiment and thermal test, including the vibrations, shocks, accelerations and impact test during the launch.

\subsection{Electronic design}

\begin{figure}[!ht]
   \centering
   \includegraphics[width=0.8\textwidth, angle=0]{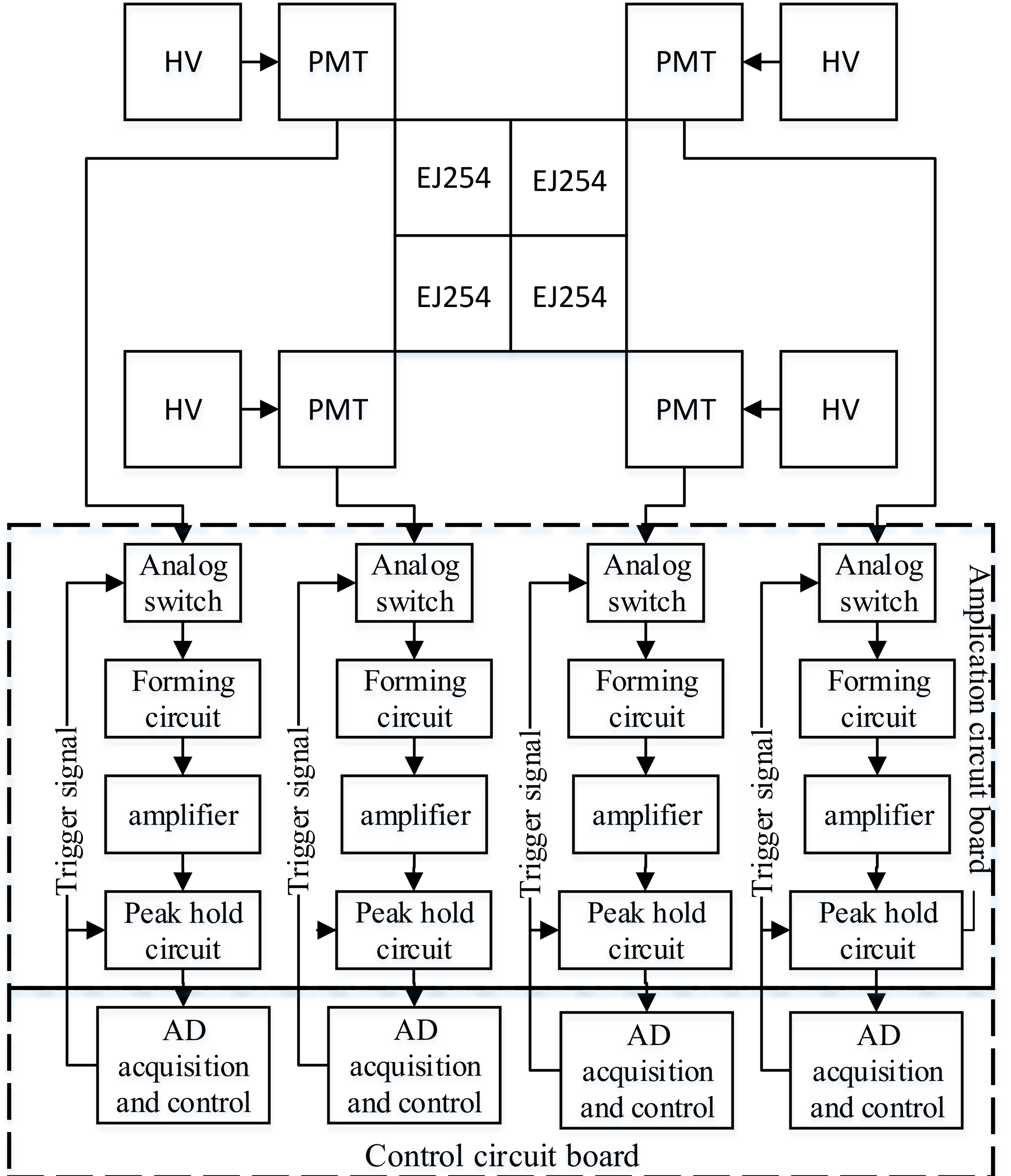}
   \caption{the circuit diagram of neutron detector}
   \label{fig2}
   \end{figure}

The readout electronics of the NUD consists of four signal channels provided in one data processing board. Each channel contains a fast pre-amplifier, a gating circuit(GC), a shaping circuit (SC) and a main amplifier with peak holding
chip (PHC). Figure~\ref{fig2} shows the overall circuit diagram. The GC and PHC are controlled by the data control unit of the DAMPE satellite. The GC is designed to prevent any early signal entering the SC, and is switched-on 1.6 $\mu$s after the triggering signal produced by BGO. Then the delayed neutron signal could be shaped and amplified to the PHC. After the ADC finishes the acquisition of all four signals, a release signal will be sent to the PHC and GC to shut off the signal channel and wait for the next trigger[\citealt{ChangJ2017}].

A delay time of 2.5$\mu$s after the trigger could basically prevent the ionization signals from secondary charged particles, as the shower development would typically finish within 1$\mu$s. Afterward the analog switch will turn on and  the signals from PMT will be send to the shaping circuit and the amplifier. At the same time the peak detection circuit will start work and catch the maximum signal voltage from amplifer. After 10$\mu$s the peak detection circuit will close the gate and hold signals, wait acquisition unit for neutron detector in data process module to sample and digitalize. The results will be packaged and stored in satellite to finish the whole acquisition procedure. Then the analog switch will be closed and the peak detection circuit will be discharged to wait a new trigger for the next acquisition.

\subsection{Structural Design}

 \begin{figure}[!ht]
   \centering
   \includegraphics[width=0.6\textwidth, angle=0]{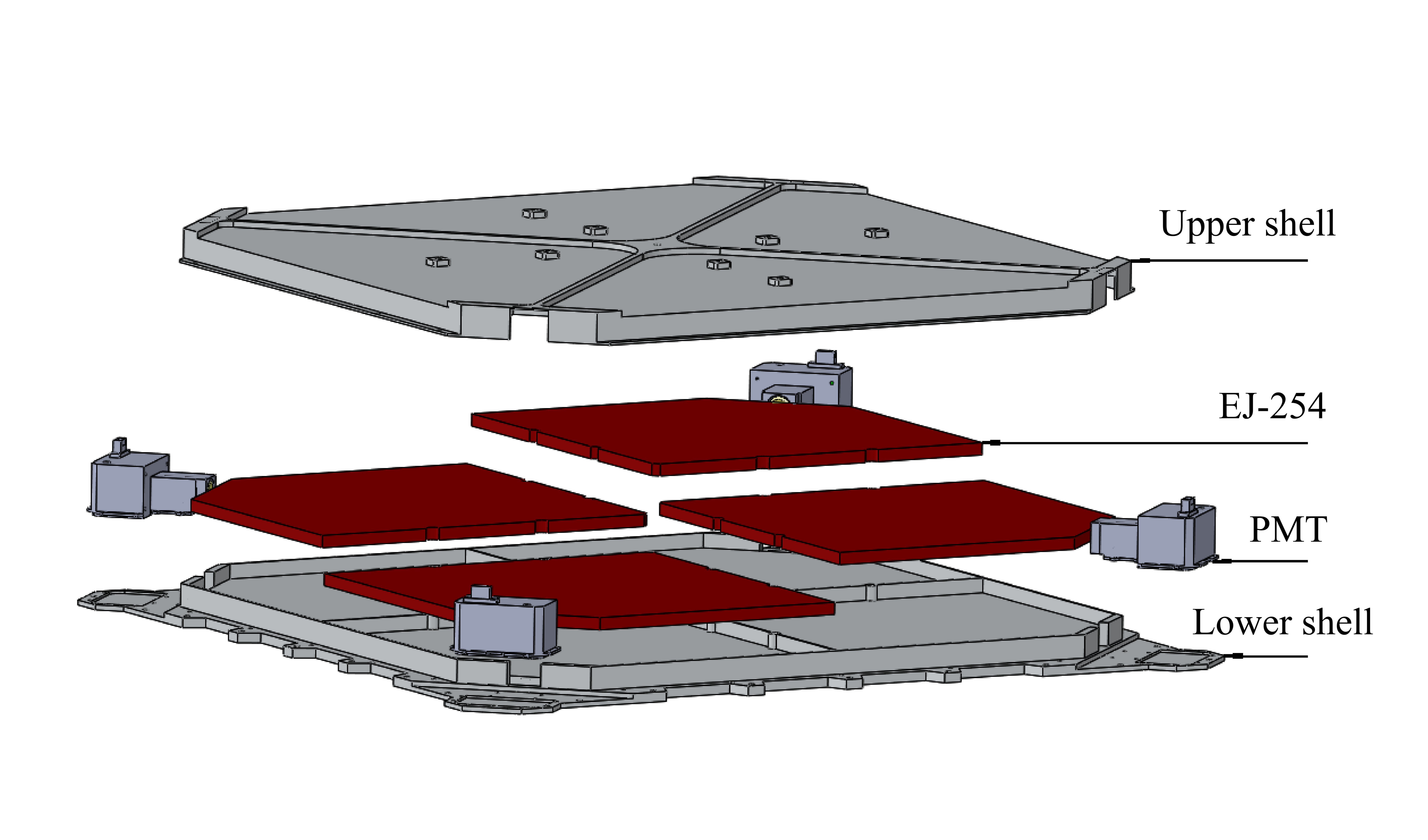}
   \caption{The structure composition of NUD}
   \label{fig3}
   \end{figure}

Four EJ-254 scintillators with the size of $300mm\times300mm\times10$mm are selected for the neutron detector. The upper and lower sides of the EJ-254 are embedded with wavelength shift fibers for optical transmission in order to reduce the fluorescence attenuation and increase photon collection efficiency. Each EJ-254 is wrapped with a layer of aluminum film for photon reflection and a layer of black tape on the outside, anchored in aluminum alloy framework by silicone rubber, and readout by a PMT(Hamamatsu R5610A-01),the PMT is a 0.75inch diameter head-on, 10-dynode PMT with a maximum gain of $2\times10^{6}$, and a spectral response ranging from 300 nm to 650 nm, which is a good match to EJ-254’s 425 nm maximum emission wavelength. The space between each EJ-254 and aluminum alloy framework is 1mm on each side, and is filled with silicone rubber to relieve the vibration during the launch[\citealt{ChangJ2017}]. The supporting structure is aluminum alloy shell. The envelope size of neutron detector is $699 \times 699 \times 44$ mm$^3$ and the weight is 12 kg. Figure~\ref{fig3} shows the overall structure of neutron detector, which is fixed to the satellite with 41 M4 screws.

\section{Ground calibrations}

\begin{figure}[!ht]
   \centering
   \includegraphics[width=0.6\textwidth, angle=0]{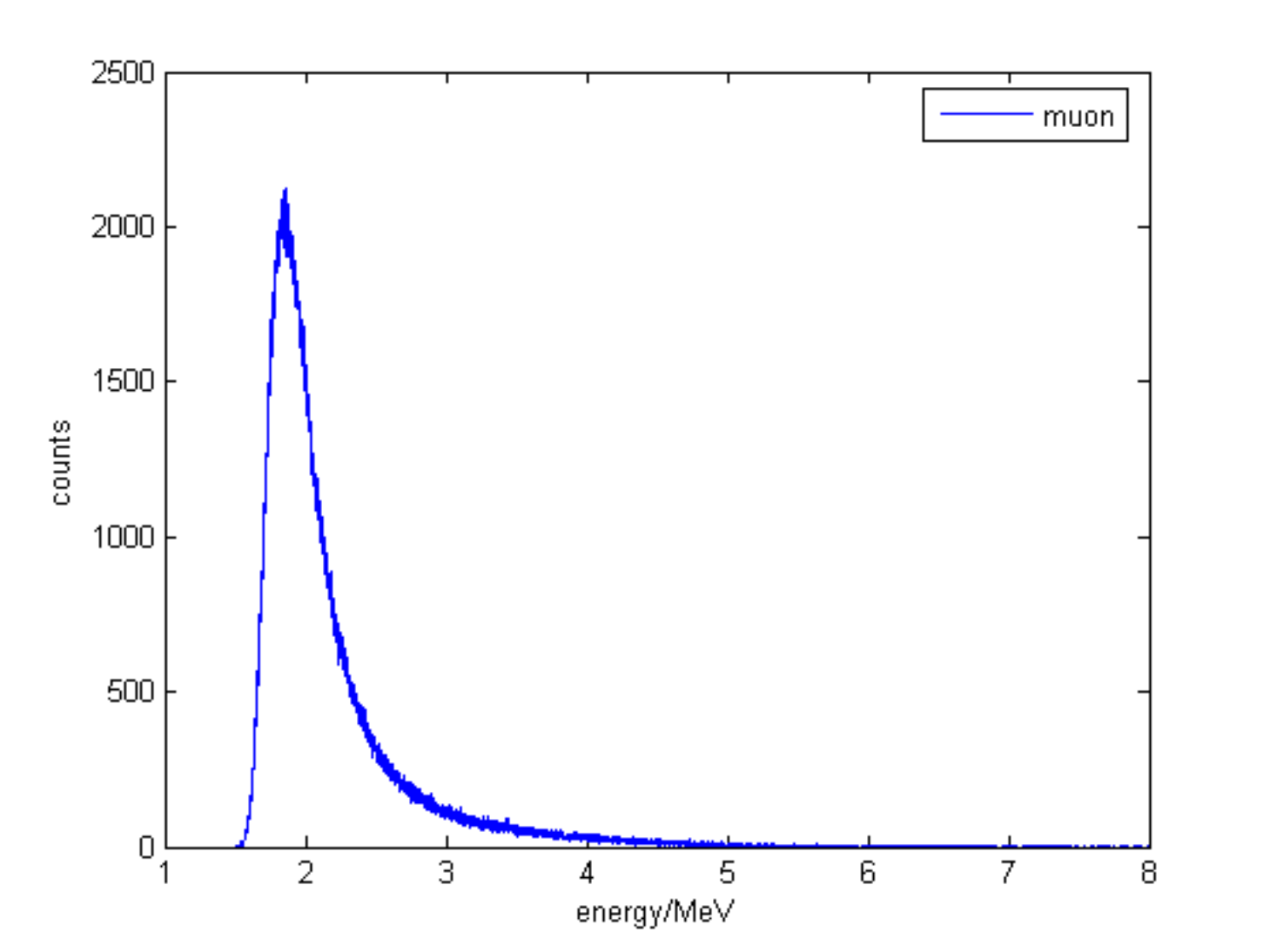}
   \caption{Simulation of energy deposition of muon in NUD}
   \label{fig4}
   \end{figure}

The dynamic range, energy resolution, the difference of each detection unit and the light attenuation characteristics of different hit positions of the neutron detector need to be calibrated on the ground. In the calibration scheme, we use muon as a source for performance tests. The deposition energy of muon in neutron detector is the minimum ionization energy loss. Figure~\ref{fig4} shows the energy deposition spectrum in neutron detector, and its peak position corresponds to 1.83 MeV.

\begin{figure}[!ht]
   \centering
   \includegraphics[width=0.5\textwidth, angle=0]{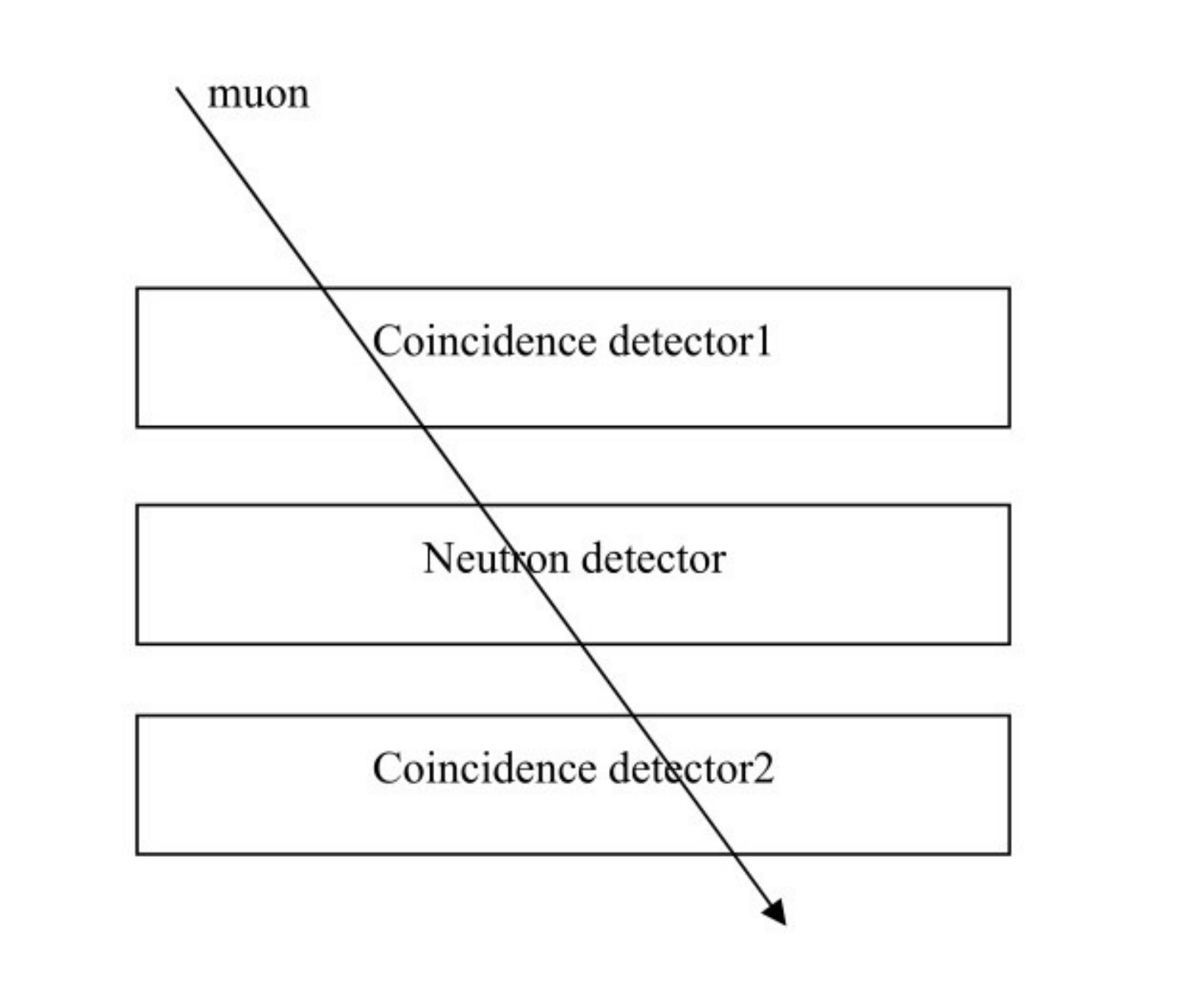}
   \caption{the schematic diagram of calibration test}
   \label{fig5}
   \end{figure}

In the calibration experiment, we use muon as source to carry out the calibration experiment and use two bigger identical plastic scintillator detectors as coincidence detectors. The detection area of plastic scintillator detector covers the area of neutron detector, Two plastic scintillator detectors are located above and below the neutron detector, Figure~\ref{fig5} shows the schematic diagram of our calibration test.

When the two plastic scintillators have hit signals at the same time, they will generate hit signals through logical AND and neutron detector starts to measure the energy deposition of muon in the neutron detector. By collecting the energy deposition spectrum for a long time, the deposition energy spectrum of muon in the neutron detector can be obtained. In practice, the output signal of the plastic scintillator is sent to the discriminator, the threshold value is set for the discriminator to filter out the interference of the electronic noise, the trigger signal will be generated by logical operation of the signal from the discriminator. Then the counter records the effective trigger times, while  the energy spectrum of muon in neutron detector will be collected.

\begin{figure}[!ht]
\centering
\subfigure[Detection unit I]{
\includegraphics[width=7cm,angle=0]{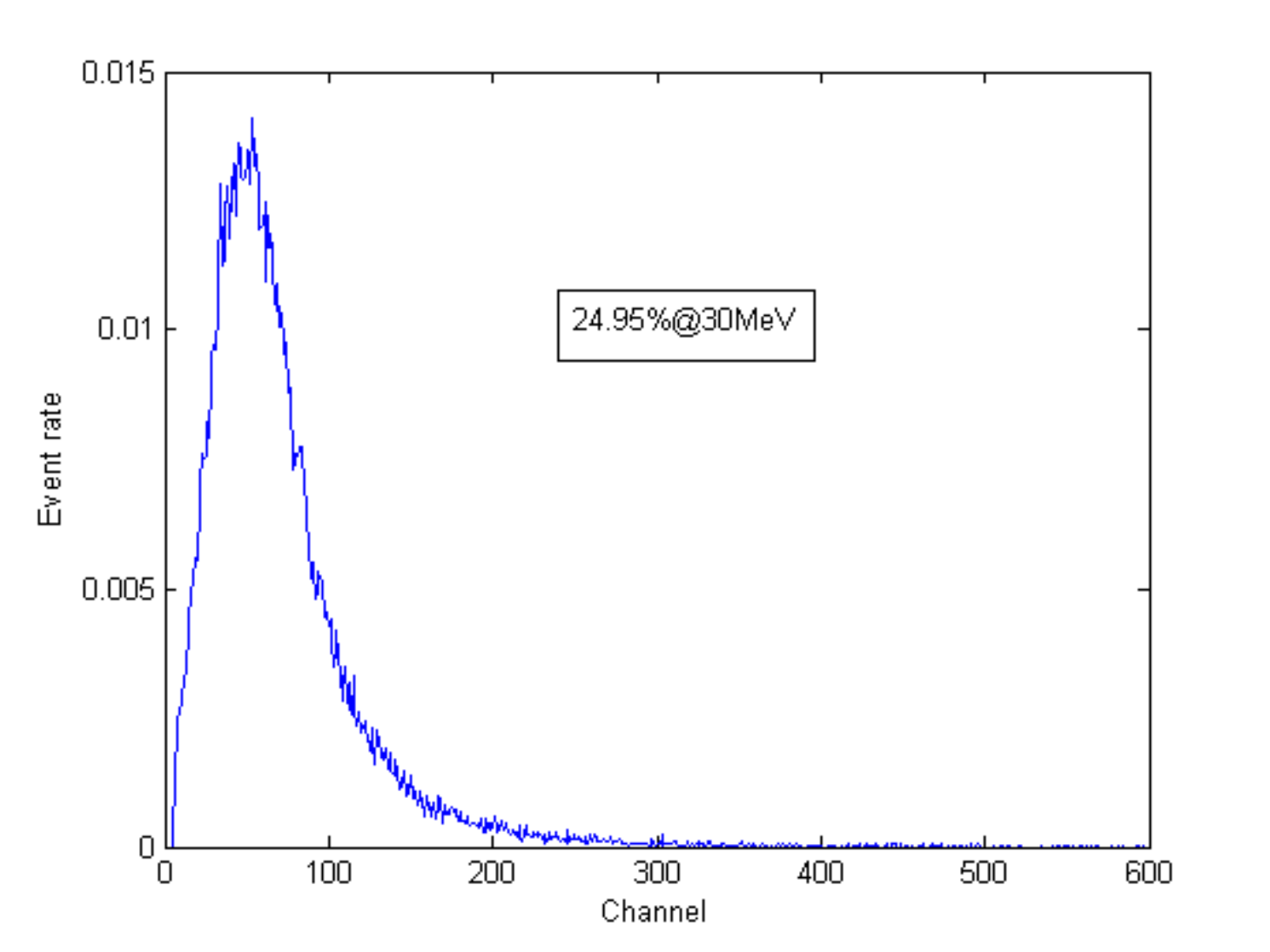}}
\subfigure[Detection unit II]{
\includegraphics[width=7cm,angle=0]{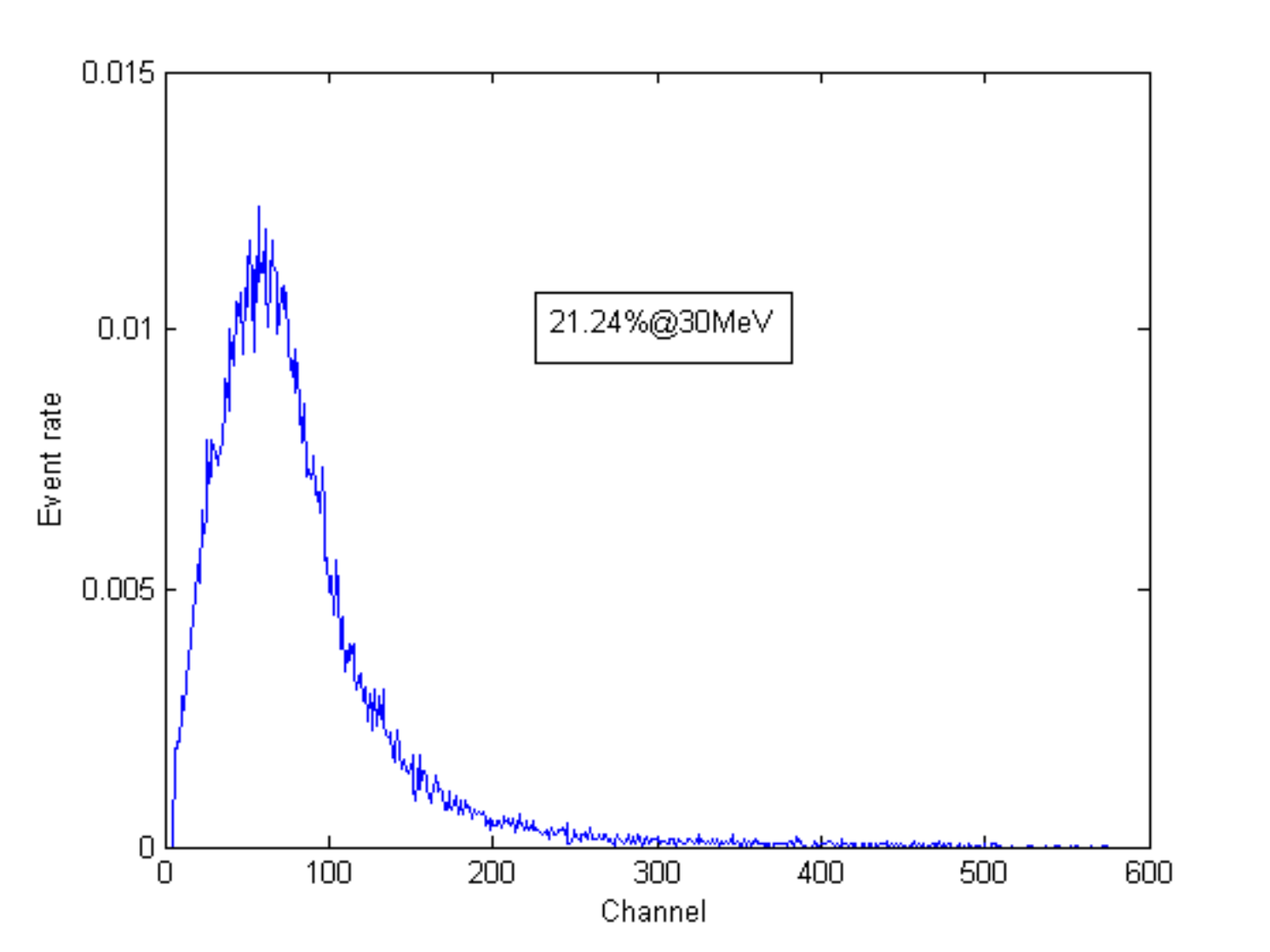}}
\subfigure[Detection unit III]{
\includegraphics[width=7cm,angle=0]{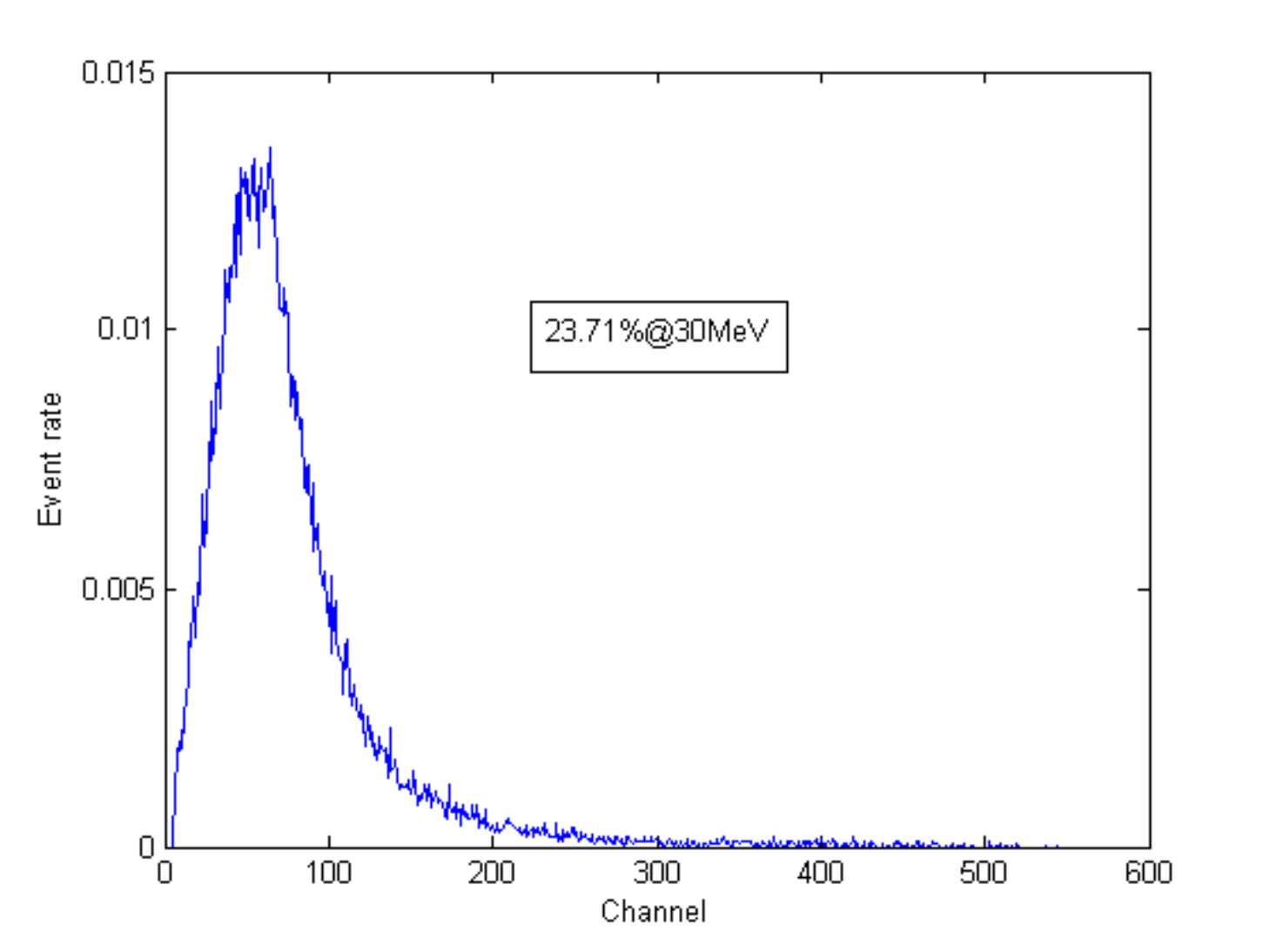}}
\subfigure[Detection unit IV]{
\includegraphics[width=7cm,angle=0]{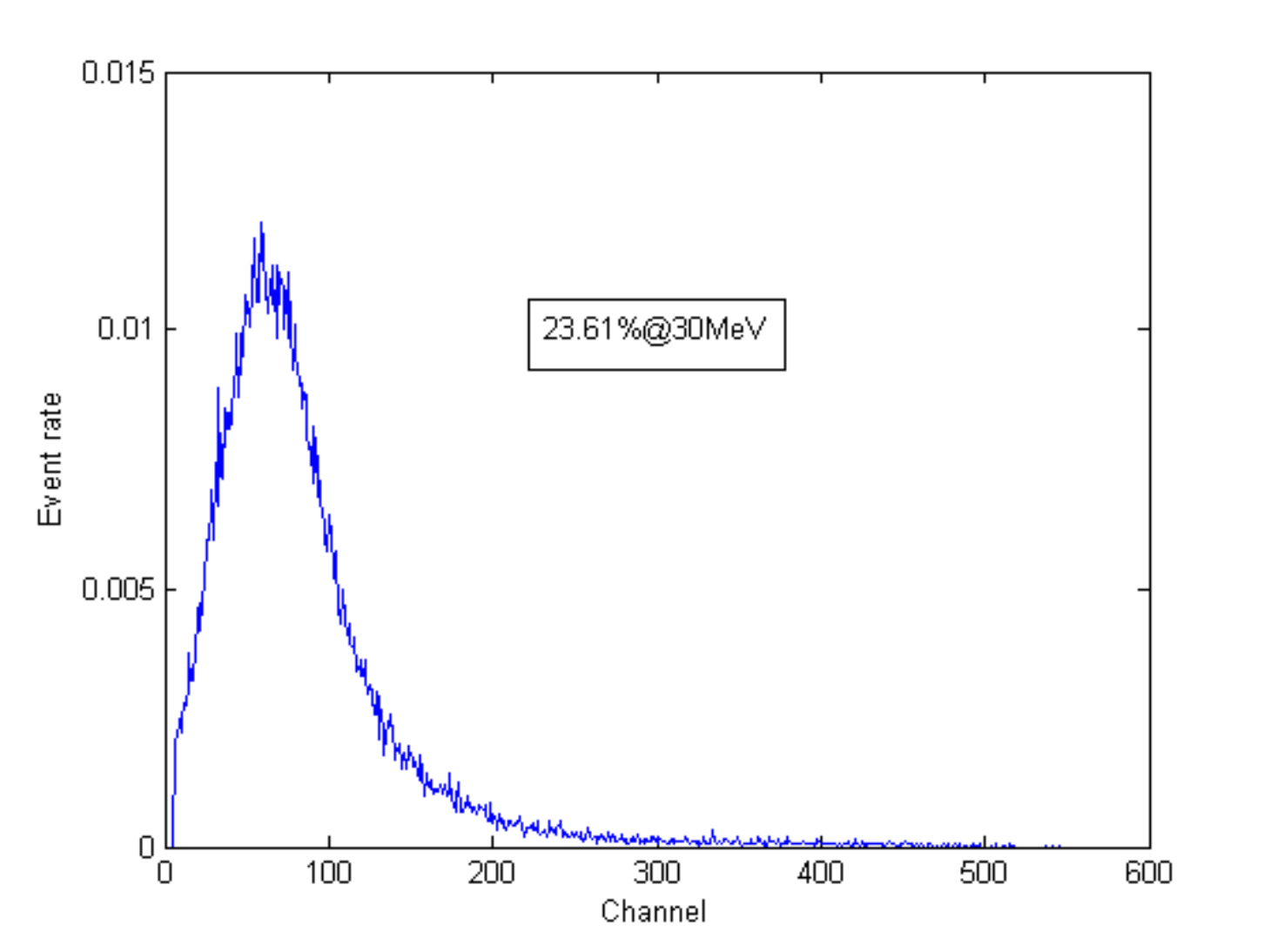}}
\caption{Energy resolutions of four detector units}
\label{fig6}
\end{figure}

\begin{tiny}[!ht]
\begin{table}
\caption[]{The difference of four channels}
\centering
\setlength{\tabcolsep}{2pt}
 \begin{tabular}{|c|c|c|c|c|c|}
  \hline
  Project & detection unit I  & detection unit II  & detection unit III  & detection unit IV & Maximun relative error \\
  \hline
  Energy resolution & 24.95\%   & 21.24\%   & 23.71\%  &  23.61\%   &  14.87\%  \\
  \hline
  Dynamic range & 0.31-60.94MeV   & 0.34-67.58MeV   & 0.38-73.92MeV  &  0.31-60.5MeV   &  18.15\%  \\
  \hline
  Detection efficiency & 97.78\%   & 99.06\%   & 100\%  &  98.8\%   &  2.22\%  \\
  \hline
\end{tabular}
\label{tab1}
\end{table}
\end{tiny}

The energy resolutions of four detection units are calibrated as 24.95\%, 21.24\%, 23.71\% and 23.61\% at 30 MeV, for unit I, II, III and IV, respectively, as  shown in Figure~\ref{fig6}. The energy resolutions of the four detection units can well meet the requirement of less than 25\%. Under 840V high voltage power supply, the dynamic range of neutron detector is measured as shown in table~\ref{tab1}. The dynamic range of detector can fully meet the detection requirements.

\begin{figure}[!ht]
\centering
\subfigure[Detection unit I]{
\includegraphics[width=7cm,angle=0]{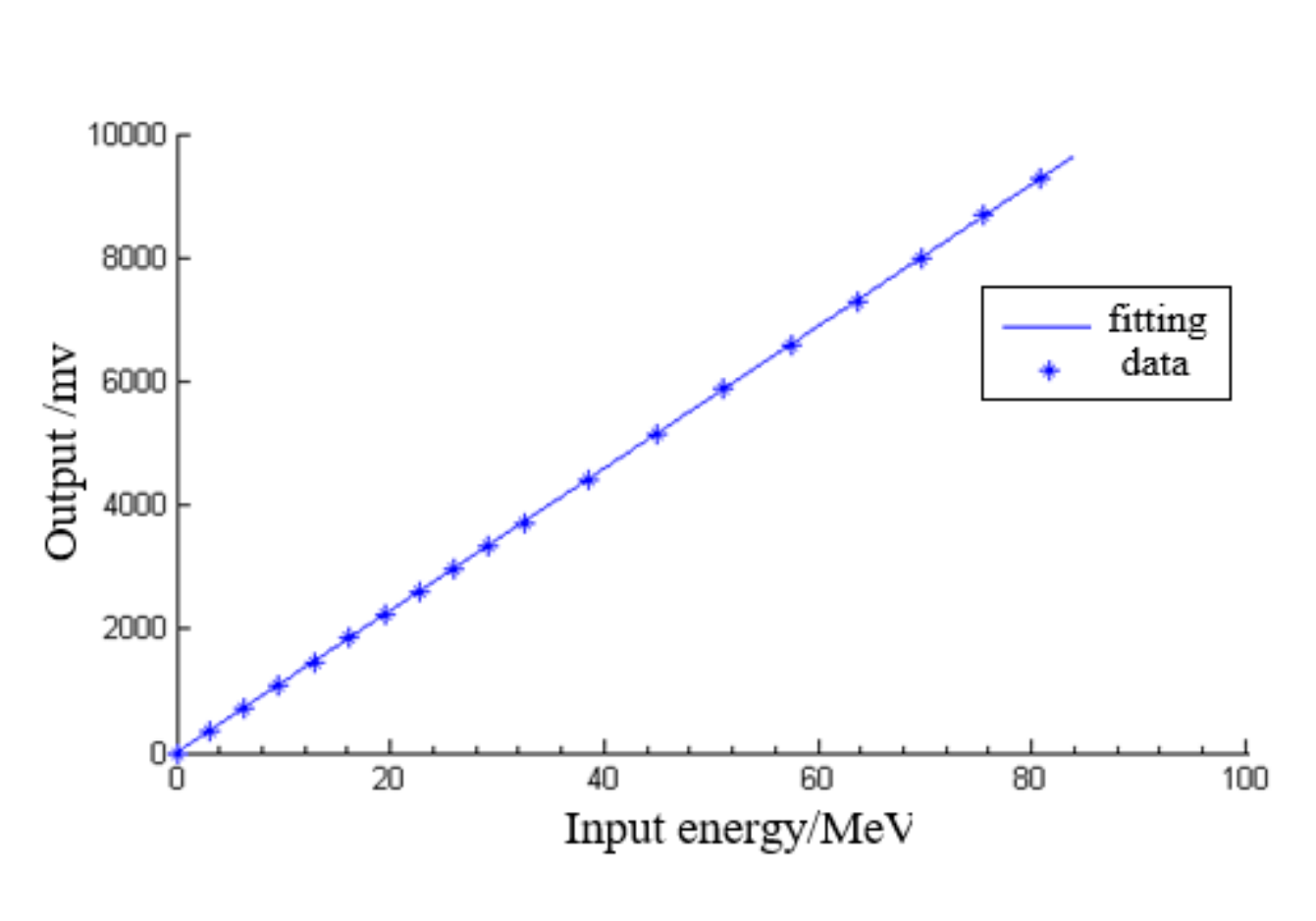}}
\subfigure[Detection unit II]{
\includegraphics[width=7cm,angle=0]{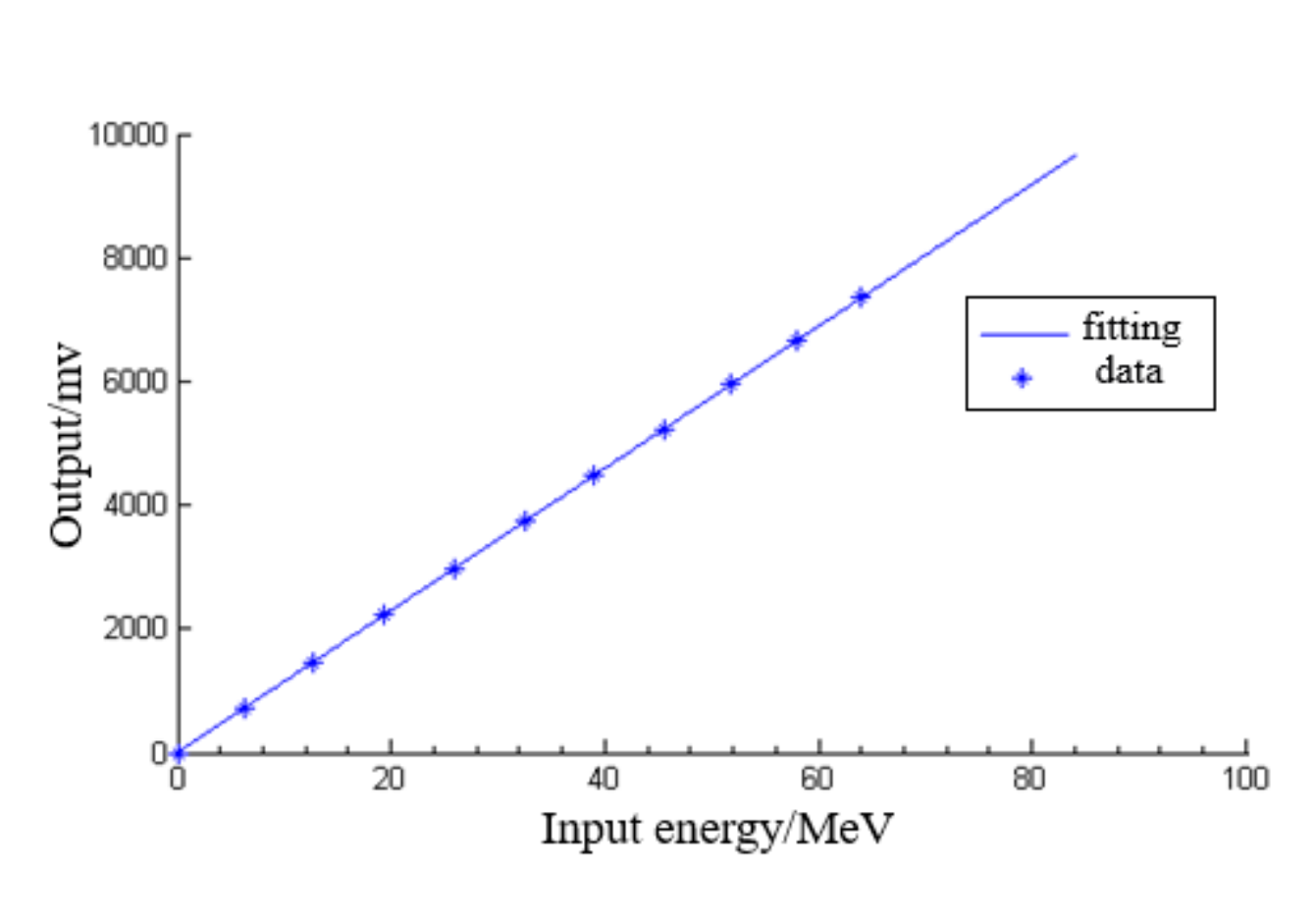}}
\subfigure[Detection unit III]{
\includegraphics[width=7cm,angle=0]{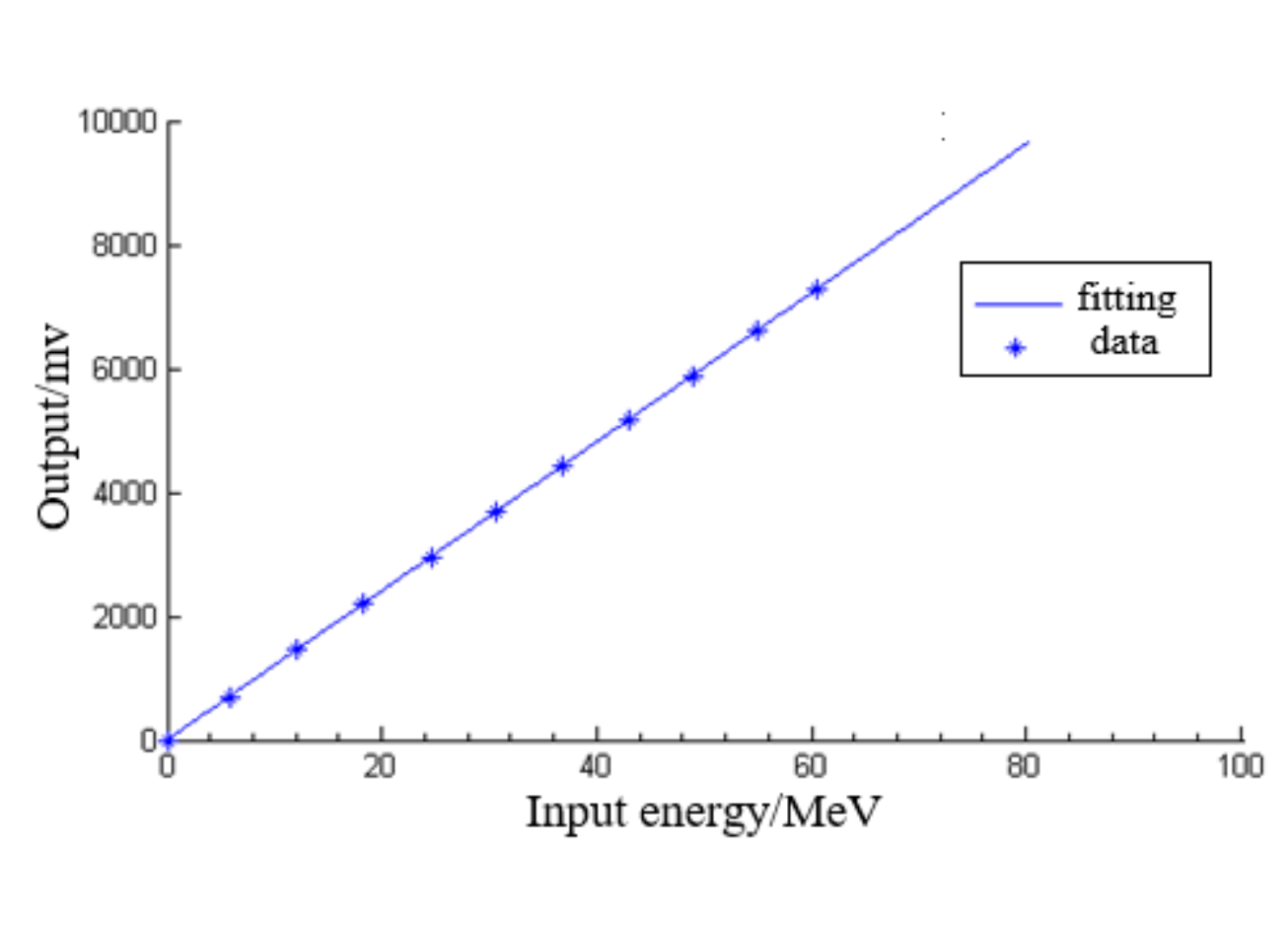}}
\subfigure[Detection unit IV]{
\includegraphics[width=7cm,angle=0]{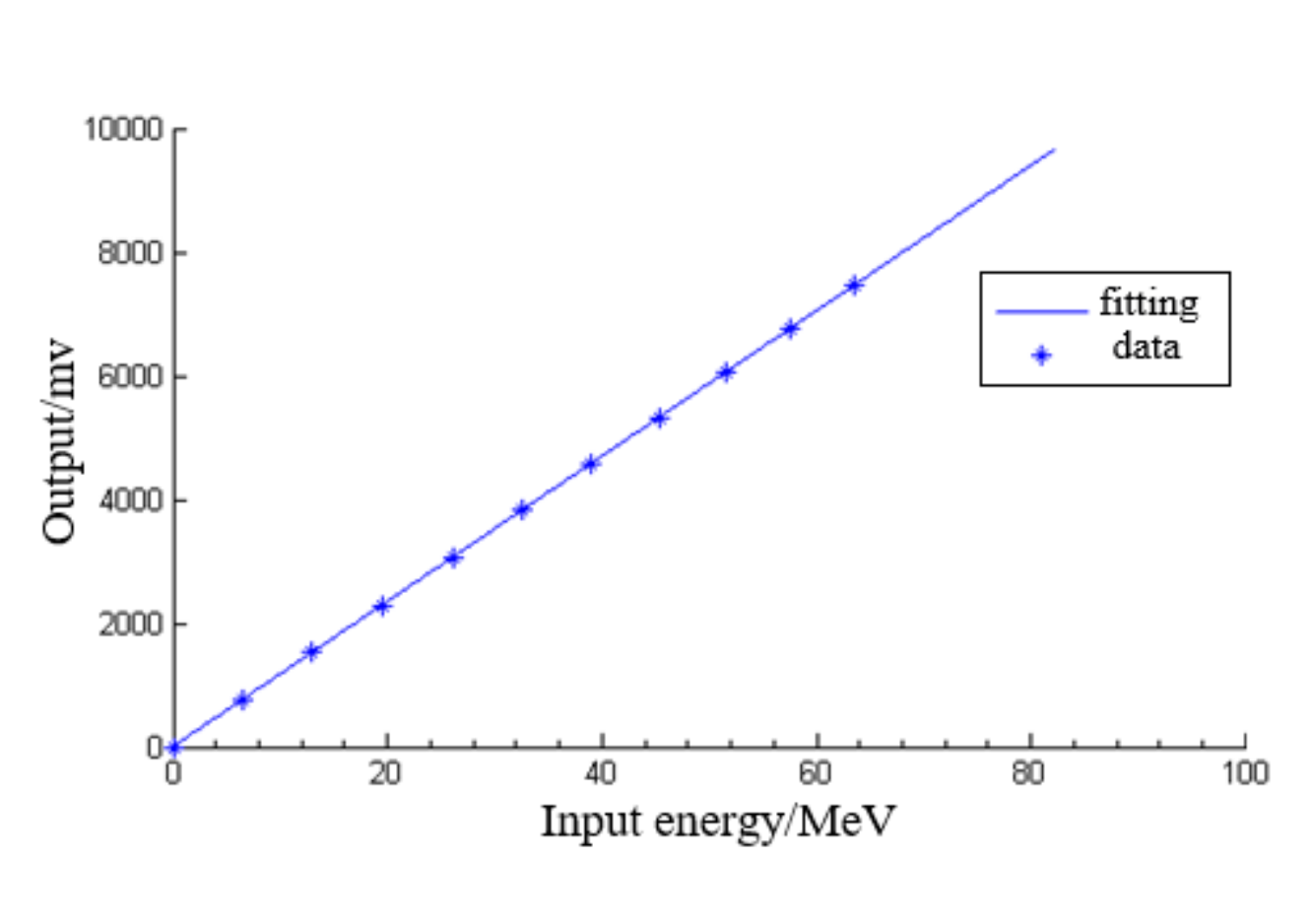}}
\caption{Dynamic range of four detector units}
\label{fig7}
\end{figure}

 \begin{figure}[!ht]
   \centering
   \includegraphics[width=0.5\textwidth, angle=0]{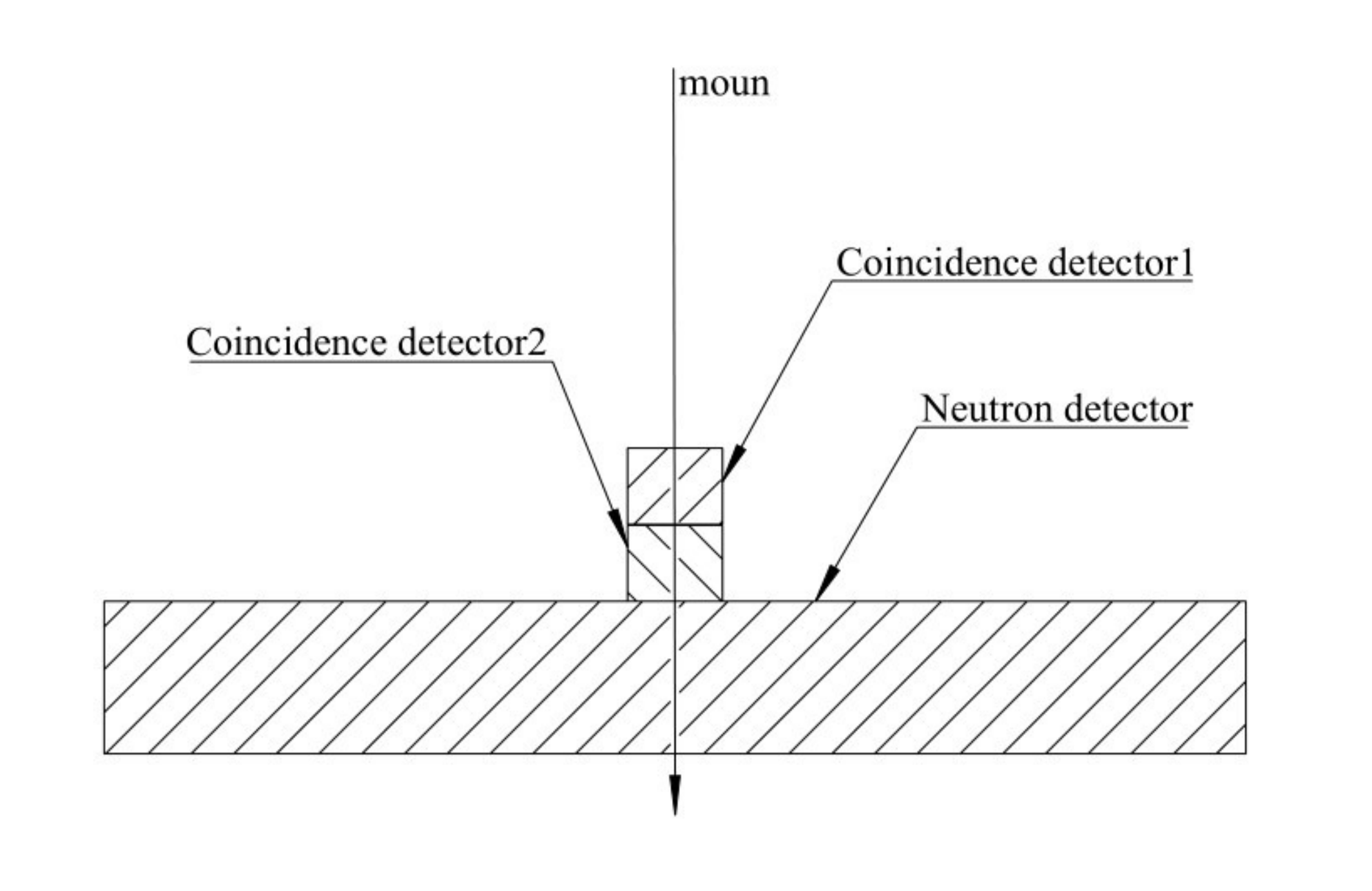}
   \caption{Schematic diagram of test scheme}
   \label{fig8}
   \end{figure}

To calibrate the hit output performance of the neutron detector, we use two smaller coincidence detectors placed in the central area of the neutron detector. The position relationship between the neutron detector and the small coincidence detector 1 and 2 is shown in the Figure~\ref{fig8}.

The detection efficiency of the central area is the ratio of the acquisition energy spectrum count and the trigger count. The four channels of the neutron detector use the above steps to acquire and count 3600 seconds respectively. The results are shown in the table~\ref{tab2}.The table~\ref{tab2} shows that the detection efficiency of the four detection units is higher than 96\%.
There are differences in energy resolution, dynamic range and muon detection efficiency of four detection units of neutron detector. The differences are mainly caused by encapsulation, detection materials and other factors. The differences among the four detection units are inevitable. We compare the energy spectrum of the four detection units, and use the relative error to quantitatively describe their differences through their energy resolution, energy dynamic range and muon detection efficiency, as shown in table~\ref{tab1}. Even though there are small differences among the four channels, it will not affect the normal operation of the neutron detector.

\begin{tiny}[!ht]
\begin{table}
\caption[]{Detection efficiency of central area}
\centering
\setlength{\tabcolsep}{2pt}
 \begin{tabular}{|c|c|c|c|c|c|}
  \hline
  Detection unit & Time (s) & Trigger count & Detection count  & Trigger frequency (counts/min) & Detection efficiency \\
  \hline
  Unit I & 3600 & 90 & 88 & 1.50 &  97.78\%  \\
  \hline
  Unit II & 3600 & 107 & 106 &  1.78 & 99.06\%  \\
  \hline
  Unit III & 3600 & 85 & 85 & 1.41 & 100\%  \\
  \hline
  Unit IV & 3600 & 83 & 82 & 1.38 & 98.80\%  \\
  \hline
\end{tabular}
\label{tab2}
\end{table}
\end{tiny}

\section{On-orbit performance}

To carefully estimate the performance of DAMPE detector, including NUD, we have developed an extensive Monte Carlo (MC) simulation software based upon the latest GEANT4 toolkit [\citealt{Allison2016}]. The hadronic interaction model FTFP\_BERT\_HP, which includes high-precise neutron cross-section data, has been chosen for high energy proton and electron simulations. A digitization algorithm for NUD has been developed by simulating the procedure of A/D conversion in the electronics, thereby converting energy deposits into digital signals which have same format as the raw flight data. In the digitization procedure, the same delay time of 2.5$\mu$s is applied to keep up with the flight data.

 \begin{figure}[!ht]
   \centering
   \includegraphics[width=0.6\textwidth, angle=0]{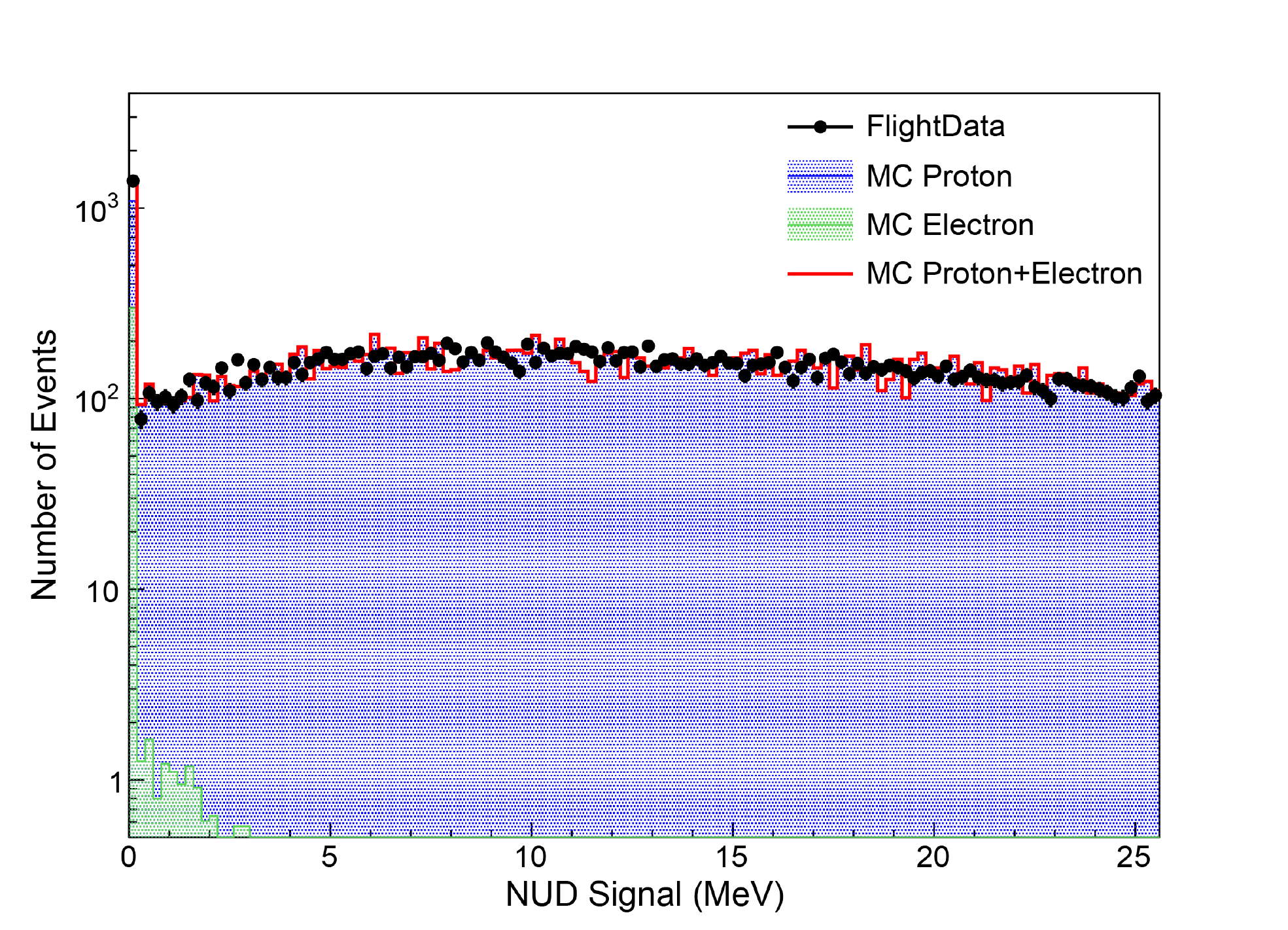}
   \caption{Comparisons of the energy signal in NUD between flight data and simulated data with energy deposit between 1TeV and 5TeV. The black points represent flight data. For MC simulated data, the blue, green and red histograms represent the protons, the electrons and their sum, respectively.}
   \label{fig9}
   \end{figure}

To validate the MC simulation, we select high energy protons and electrons from flight data based on the preselection criteria applied in DAMPE proton spectrum analysis [\citealt{AnQ2019}]. The Figure~\ref{fig9} shows the comparisons of the energy signal in NUD between flight data and MC simulated data, for protons and electrons with energy deposit in the BGO-ECAL between 1TeV and 5TeV. The NUD signals of protons show dramatic difference with the ones of electrons, which suggests that NUD has good capability for e/p discrimination.The good agreement between flight data and simulated data indicate that the MC simulation is reliable for further study.

 \begin{figure}[!ht]
   \centering
   \includegraphics[width=0.6\textwidth, angle=0]{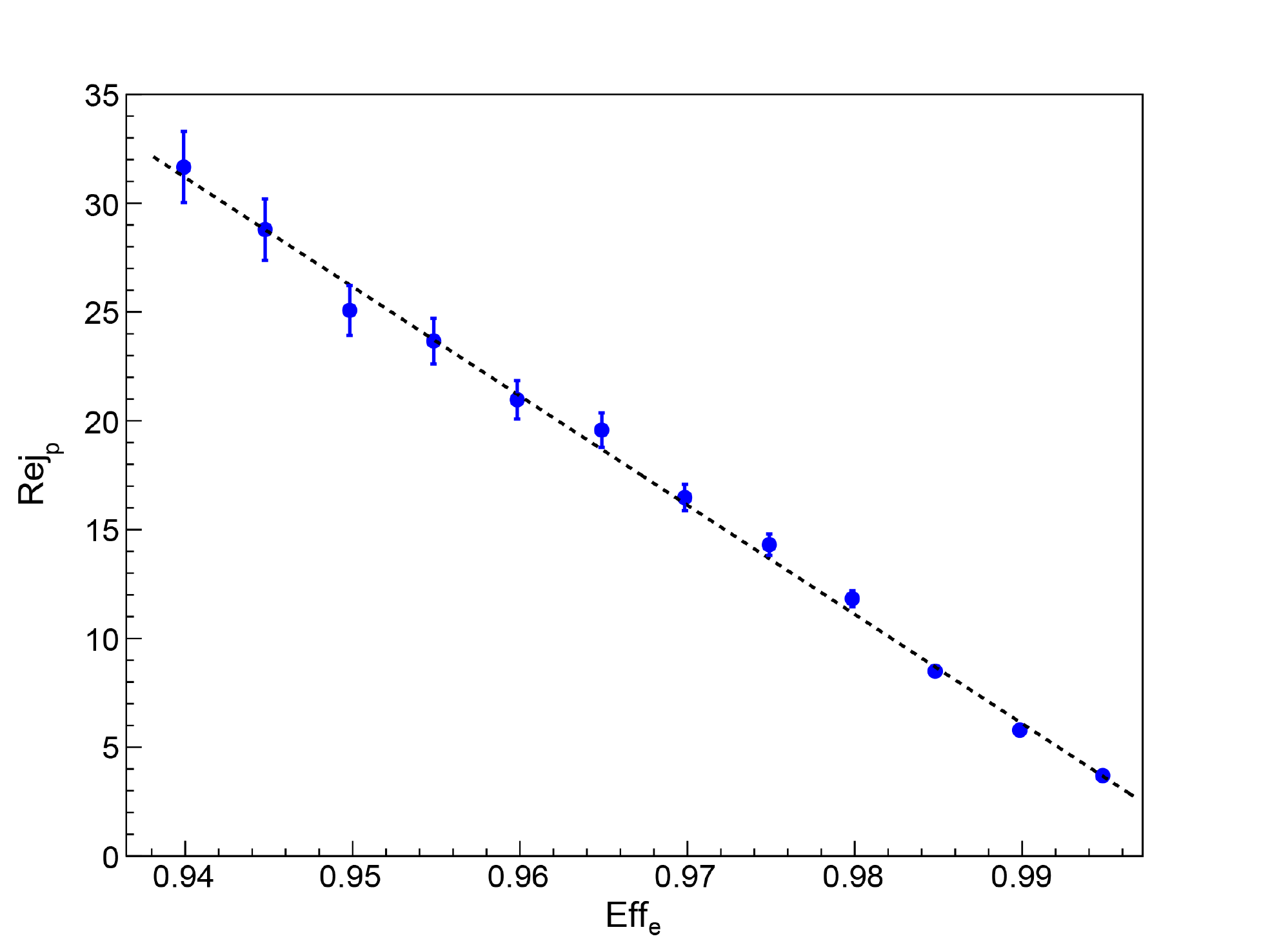}
   \caption{The rejection power of NUD for energy deposit between 1TeV and 5TeV.}
   \label{fig10}
   \end{figure}

The e/p discrimination capability of NUD is studied with the simulated protons and electrons. To characterize the e/p discrimination power, we define electron efficiency and proton rejection as following:
\begin{equation}
Eff_{e}=\frac{Ne_{sel}}{Ne_{acc}}; \; Rej_{p}=\frac{Np_{acc}}{Np_{sel}}
\label{eq-epRej}
\end{equation}
where $Ne_{acc}$ and $Ne_{sel}$ represent the number of accepted electron and the number of electron selected by NUD,  $Np_{acc}$ and $Np_{sel}$ represent the number of accepted proton and the number of proton mistaken as electron by NUD. Figure~\ref{fig10} shows the relationship curve between the electron efficiency and the proton rejection.  When the electron efficiency is 95\% the proton rejection power is $\sim$25, which indicates that the NUD can distinguish electrons and protons effectively under a satisfying electron acceptability.

 \begin{figure}[!ht]
   \centering
   \includegraphics[width=0.48\textwidth, angle=0]{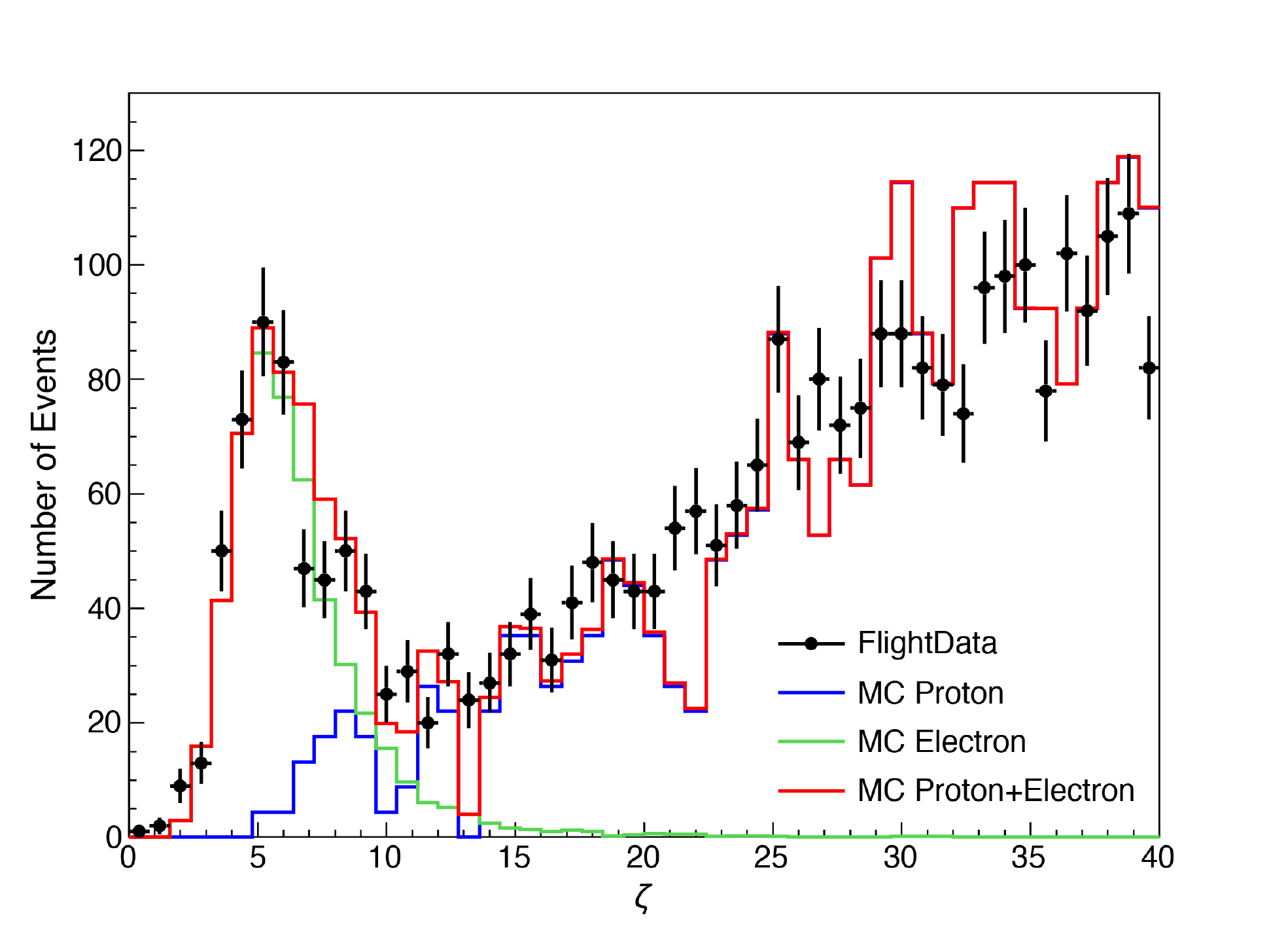}
   \includegraphics[width=0.48\textwidth, angle=0]{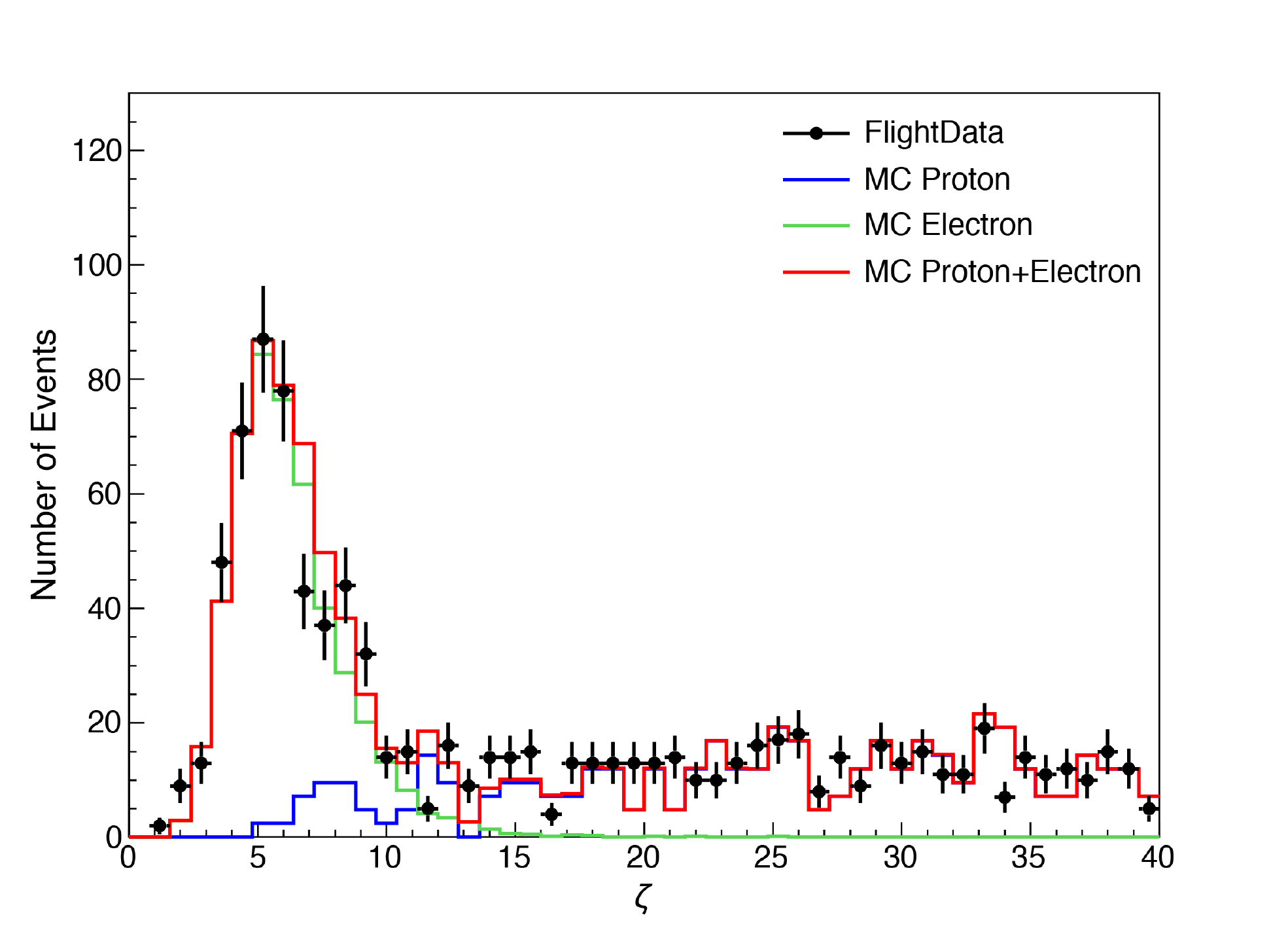}
   \caption{The $\zeta$ distributions for on-orbit events with energy deposit from 1TeV to 5TeV before ($left$) and after ($right$) NUD selection. The black points represent the flight data. For MC simulated data, the blue, green and red histograms represent the protons, the electrons and their sum, respectively.}
   \label{fig11}
   \end{figure}

With the help of shower development information in BGO-ECAL, we can further study e/p discrimination capability of NUD with on-orbit flight data. In the calorimeter alone based electron analysis, a dimensionless variable, $\zeta$, is defined to evaluate the e/p discrimination capability of BGO-ECAL [\citealt{Ambrosi2017}]. The e/p discrimination capability of NUD is evaluated by comparing the residual proton contamination in electron candidates before and after applying the NUD selection of $Eff_{e}=95\%$. The $\zeta$ distributions for on-orbit events with energy deposit in the BGO-ECAL from 1TeV to 5TeV before (left) and after (right) NUD selection are shown in Figure~\ref{fig11}.  The MC simulated data and the flight data are in good agreement with each other for both of left and right panels. By comparing the right panel to the light one, we can see that the proton events has been effectively rejected. With a  $\zeta$ selection cut for the electron detection efficiency of 90\%, the proton contamination ratio reduces by a factor of $\sim50\%$ after applying the NUD selection, which indicates that the NUD provides independent proton rejection power by a factor of $\sim$2 on the basis of BGO-ECAL.

\section{Conclusions}
The neutron detector is an essential sub-payload of the DAMPE satellite. The neutron detector is designed to detect the signals of secondary neutrons produced by the hadronic shower in the BGO-ECAL, thereby providing an important e/p separation power in high energy range. The detailed calibrations show that the neutron detector is well designed and assembled to meet the requirements. Based on the MC simulations, we obtained a proton rejection power of $\sim$25 under a electron efficiency of 95\% for the neutron detector in TeV energy range. The on-orbit performance shows that the neutron detector achieves significant proton rejection power independently from the BGO-ECAL. Thereby, the neutron detector plays an essential role for TeV electron/proton discrimination of DAMPE.

\begin{acknowledgements}
 In China, this work is supported in part by National Key Research and Development Program of China (No. 2016YFA0400201),
National Natural Science Foundation of China (Nos. 11622327, 11273070,11673075, U1738205, U1738121, U1738207, U1531126, 11873021, 11773085,11873020),
Space Science Mission Concept Research of Strategic Priority Research Program in Space Science of Chinese Academy of Sciences (No. XDA15007114).

\end{acknowledgements}

\label{lastpage}

\end{document}